\documentclass[a4paper,9pt]{article}
\usepackage{graphicx}
\usepackage[utf8]{inputenc}
\usepackage{amssymb,amsmath,array}

\begin{document}
\title{Complex activity patterns generated by short-term synaptic plasticity}

\author{Bulcs\'u S\'andor and Claudius Gros
%
%
\vspace{.3cm}\\
%
Institute for Theoretical Physics - Goethe University Frankfurt\\
Frankfurt am Main - Germany
}

\maketitle


\begin{abstract}
%
%
Short-term synaptic plasticity (STSP) affects
the efficiency of synaptic transmission for 
persistent presynaptic activities. We consider 
attractor neural networks, for which the
attractors are given, in the absence of STSP,
by cell assemblies of excitatory cliques. We 
show that STSP may transform these attracting states 
into attractor relics, inducing ongoing transient-state 
dynamics in terms of sequences of transiently 
activated cell assemblies, the former attractors. 
Subsequent cell assemblies may be both disjoint or 
partially overlapping. It may hence be possible to use
the resulting dynamics for the generation of motor 
control sequences.
\end{abstract}

%
%


\section{Introduction}

Changes in the transmission properties of synapses 
lead to a modulation of information processing. Synaptic 
plasticity, which is present on many time scales,
is responsible, for learning and computational 
processes \cite{cortes2013short}, as well as for memory 
consolidation \cite{mongillo2008synaptic} and motor 
pattern generation \cite{martin2016closed}.
Short-term synaptic plasticity (STSP) contributes, in 
this context, to the regulation of brain networks on 
time scales ranging typically from tens of milliseconds 
to seconds \cite{gupta2000organizing}. It is affected,
in the standard Tsodyks-Markram model 
\cite{tsodyks1998neural, mongillo2008synaptic, 
cortes2013short}, by two factors: a reservoir $\varphi(t)$
of neurotransmitter vesicles and 
the Ca-concentration, $u(t)$, which in turn influences
the release probability of vesicles. The synaptic efficiency
is then proportional to the number of released neurotransmitters 
per incoming presynaptic spike, which is, in turn, proportional 
to both $u$ and $\varphi$.

\section{Methods and results}

We consider a modified version of the original 
Tsodyks-Markram model for presynaptic plasticity:
%
\begin{equation}
\begin{aligned}
\dot{u} &= {\left(U(y)-u\right)}/{T_u} 
\qquad \qquad
&U(y) &= 1+(U_{\text{max}}-1)y\nu\\
\dot{\varphi} &= {\left(\Phi(y,u)-\varphi\right)}/{T_{\varphi}} 
\qquad\qquad
&\Phi(y,u) &= 1-{uy\nu}/{U_{\text{max}}}
\end{aligned}~.
\label{eq:fdp}
\end{equation} 
%
$T_u$ and $T_{\varphi}$ set here the 
relaxation times respectively for the Ca-concentration 
$u$ and for the vesicle level $\varphi$. The target
functions $U(y)$ and $\Phi(y,u)$ depend, in addition, on 
the level $y \in [0,1]$ of the presynaptic firing rate. 
The parameter $\nu=1/0$ is used as a switch, turning
STSP on/off. The synaptic variables return 
to their baseline values, $u,\varphi\rightarrow1$, 
for both $\nu=0$ and $y=0$.

The model defined by (\ref{eq:fdp}) allows the reservoir
of neurotransmitter to fully deplete, $\varphi\rightarrow0$, 
for sustained presynaptic activity $y=1$, and for a direct 
control of the maximal Ca-concentration, $U_{\text{max}}$.
Note that STSP influences synaptic transmission only
transiently, its effects are hence fading away 
in the absence of presynaptic neural activity, viz for $y=0$. 

\begin{figure}[t]
\centering
\includegraphics[width=0.32\textwidth]
{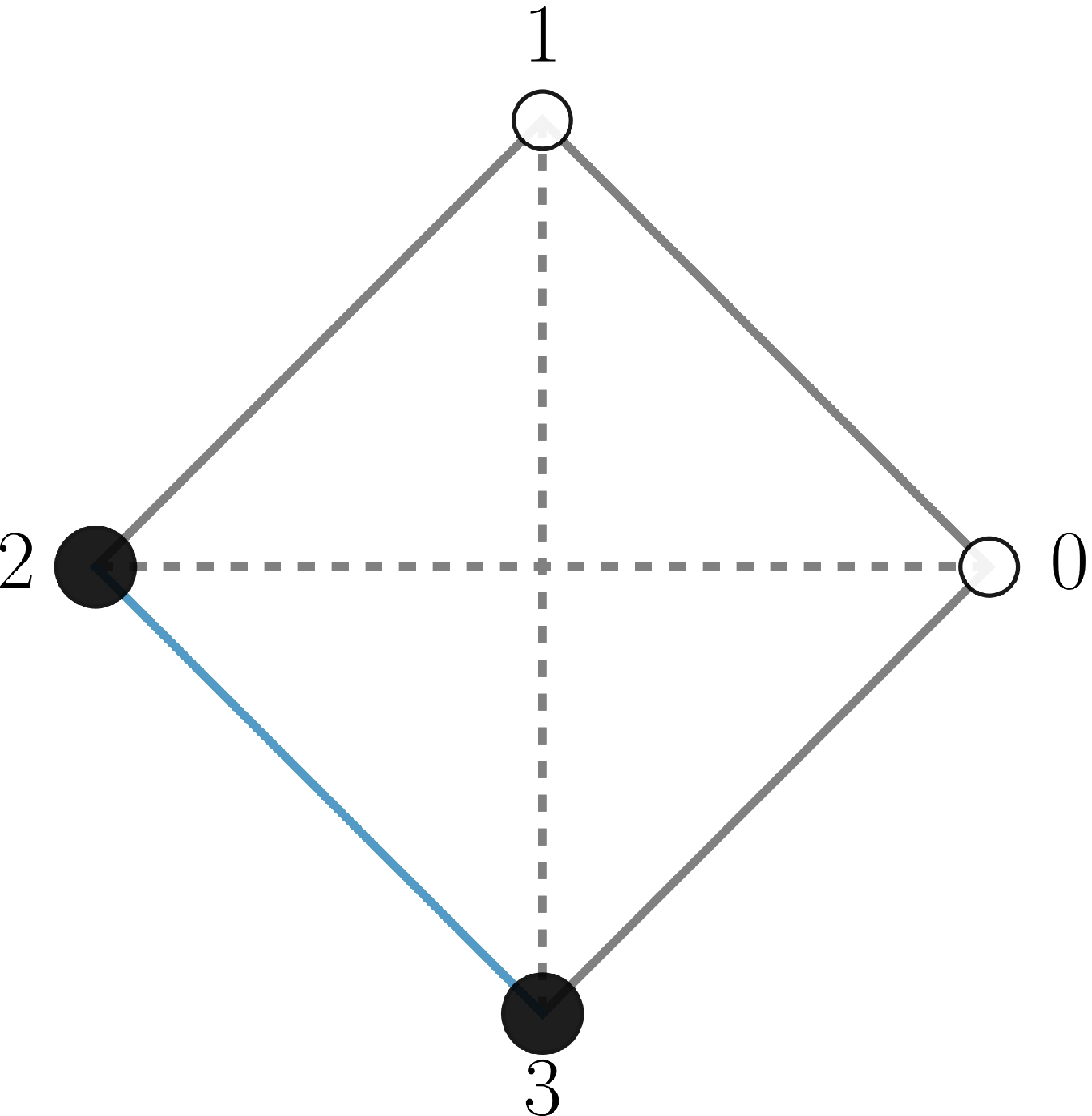}
\includegraphics[width=0.32\textwidth]
{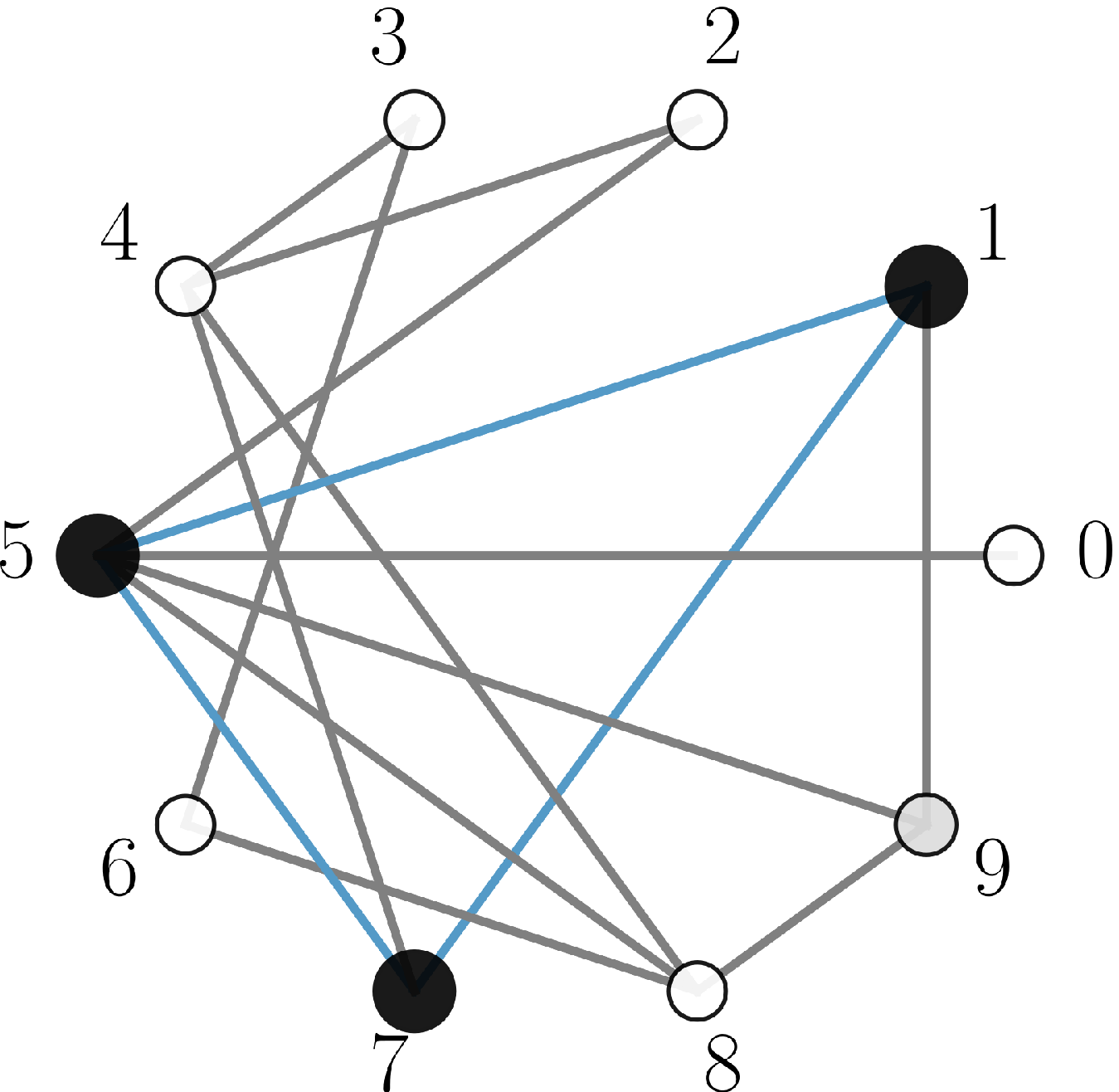}
\includegraphics[width=0.32\textwidth]
{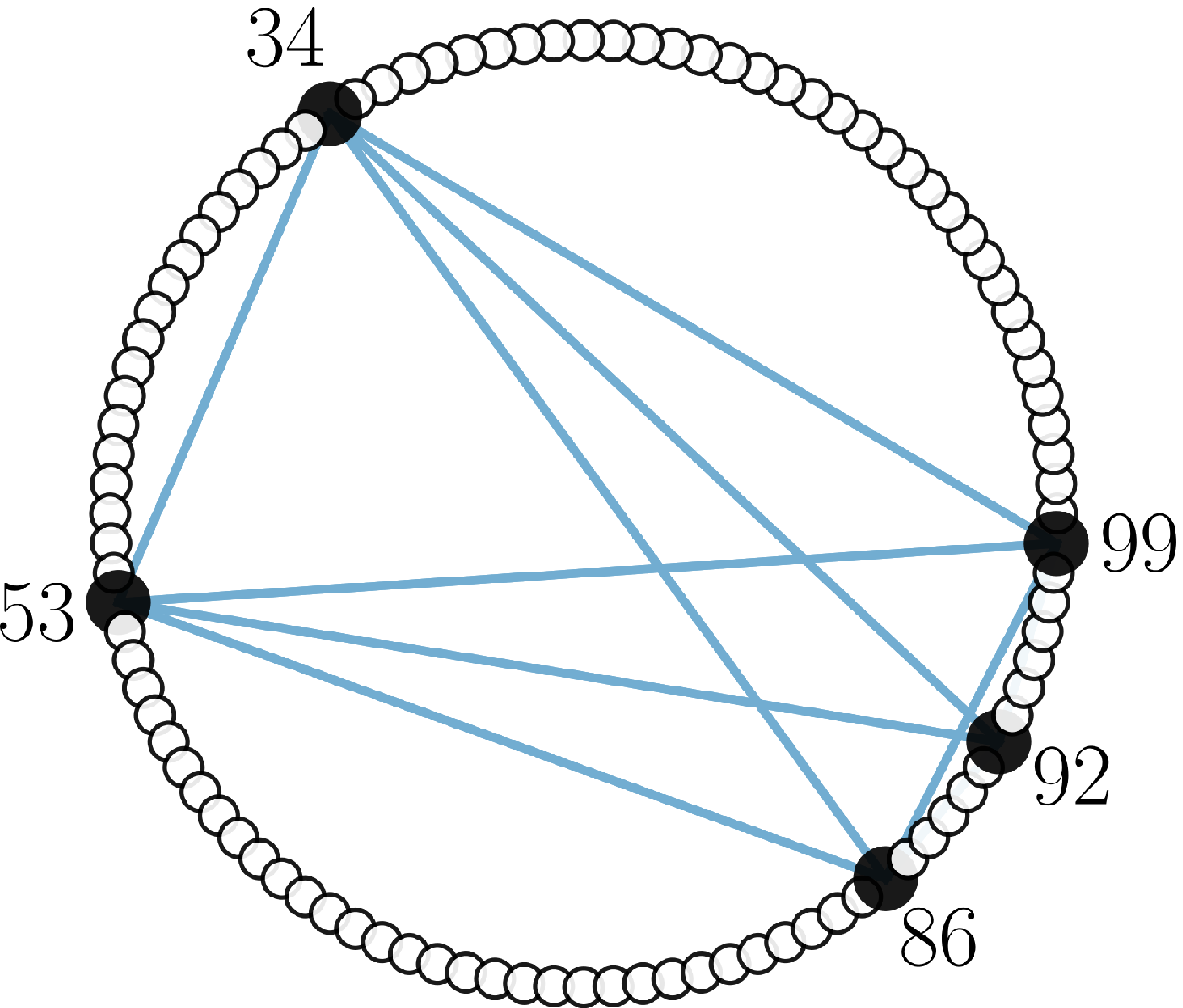}
\caption{Networks of $N=4/10/100$ neurons 
(\textit{left/middle/right}) with one active 
clique (black nodes and blue lines). Shown
are the connections (both types/only excitatory/of 
the active clique). Active/inactive nodes are 
indicated by filled/open numbered circles and
excitatory/inhibitory links by continuous/dashed lines.}
\label{fig:networks}
\end{figure}

\subsection{Transient-state dynamics with clique encoding}
We use the full depletion model, defined by Eq.~(\ref{eq:fdp}),
to generate transient-state dynamics \cite{rabinovich2008transient}, 
analogously to the method introduced in \cite{gros2007neural},
in attractor neural networks of different size. The network 
structures consist of all-to-all bidirectional 
couplings, with static excitatory connections $w_{jk}$, and 
transiently changing inhibitory connections of effective 
synaptic weights $z_{jk}u_k\varphi_k$. Excitatory and inhibitory 
couplings are mutually exclusive, $w_{jk}z_{jk}=0$, 
self-connections are excluded.

The topology of the considered networks is sketched in 
Fig.~\ref{fig:networks}. The $N=4$ site network 
has $w_{jk}=w_{kj}=w_0$ and $z_{jk}=z_{kj}=z_0$,
implying a $C_4$ rotational symmetry. The excitatory 
connections form a ring, allowing two independent 
inhibitory links. In case of the other two networks 
of $N=10$ and $N=100$ neurons, first an Erd\H{o}s-R\'enyi 
network of excitatory links is generated, with 
$w_{ij}=w_0+\eta^w_{ij}$, using a connection probability 
$p=0.3$. The $z_{ij}=z_0+\eta^z_{ij}$ are then 
complementary (not explicitly shown).
$\eta^w_{ij}$ and $\eta^z_{ij}$ are here normal
distributed random variables with respective standard 
deviations $w_{\text{std}}$ and $z_{\text{std}}$.

We consider networks of simple rate-encoding neurons, 
with a sigmoidal transfer function $y(x)=1/(1+e^{-ax})$, 
where $x$ plays the role of the membrane potential.
The dynamics of the neural activity $y_k=y(x_k)$
is defined by:
%
\begin{equation}
\label{eq:newton_neuron}
\dot{x}_j=-\Gamma x_j + \sum_{k=0}
(w_{jk}y_k+z_{jk}u_{k}\varphi_{k}y_k) + I\,,
\qquad\quad j,k\in\{0,\dots, N-1\}
\end{equation} 
%
together with the STSP rules (\ref{eq:fdp}).
$x_j$ denotes the membrane potential of the $j$-th 
neuron and $I$ a constant global input (as generated
by other brain regions). $\Gamma$ is here the 
relaxation rate of the leak term. 

The parameters are set such that the system defined
by (\ref{eq:newton_neuron}) has stable fixpoints when
STSP is absent ($\nu=0$), characterized in turn by
active cell assemblies of excitatory cliques of neurons 
(as highlighted in Fig.~\ref{fig:networks} by black nodes
and blue links). These cell assemblies realize a 
particular form of information encoding in the brain, 
called as clique encoding \cite{gros2007neural}.
STSP with $\nu=1$ destabilizes the fixpoint attractors, 
leading to transient-state dynamics corresponding to 
stable limit cycles, or chaotic attractors, in the 
full $3N$ dimensional phase space.

\begin{figure}[t]
\centering
\includegraphics[width=0.98\textwidth]
{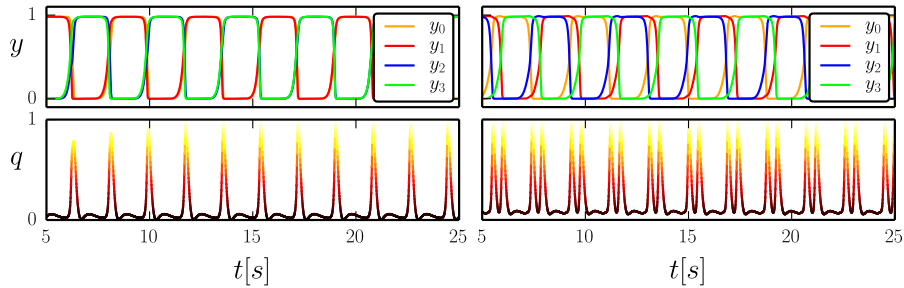}
\caption{The neural activity $y$ (\textit{top}) of the 
$N=4$ neuron symmetric network with STSP, $\nu=1$, and the 
corresponding normalized speed $q$, see Eq.~(\ref{eq:speed}), 
of the flow (\textit{bottom}).
Two examples of limit cycle solutions are shown, 
with switching (\textit{left}) and travelling wave 
(\textit{right}) dynamics. The parameters are 
$\Gamma=10\,\text{s}^{-1}$, 
$T_u=300\,\text{ms}$, $T_{\varphi}=600\,\text{ms}$, 
$w_0=40\,\text{Hz}$, $z_0=-100\,\text{Hz}$, $U_{\text{max}}=4$, 
$a=1$, $\nu=1$ and $I=0$.}
\label{fig:timeseries_N=4}
\end{figure}

As an example we present in Fig.~\ref{fig:timeseries_N=4},  
two limit cycle solutions of the $N=4$ site symmetric network.
In the first case a synchronous switching of active pairs of 
neurons is observed. This flip-flop dynamics (class I.) 
breaks the $C_4$ symmetry of the network spontaneously. 
The second, a travelling wave solution (class II.), is
characterized by an activity bump \cite{york2009recurrent}, 
which may travel clock or counter-clock wise. There are
$2$ and $4$ equivalent solutions respectively for the 
flip-flop state and for the travelling wave.

The trajectory revisits repeatedly the destabilized 
fixpoints, slowing down in their neighborhood.
To quantify this behavior, we use the normalized 
speed $q$ of the flow in the phase-space 
\cite{sussillo2013opening}, as defined by:
\begin{equation}
q=Q/Q_{\text{max}}\,,
\qquad
Q=\lvert \mathbf{f}(\mathbf{v})\rvert^2\,,
\qquad\quad
\dot{\mathbf{v}}=\mathbf{f}(\mathbf{v})\,,
\label{eq:speed}
\end{equation} 
with $\mathbf{v}=(\mathbf{x},\mathbf{u},\boldsymbol{\varphi})$ 
and with $\mathbf{f}$ denoting the right-hand side 
function of the differential equation. 
The normalization is done with respect to 
the maximal flow speed $Q_{\text{max}}$ 
of the respective limit cycle. The normalized speed
$q$ peaks, as shown in Fig.~\ref{fig:timeseries_N=4}, 
at the transition points, being close to zero when the 
cell assemblies are transiently activated. 

Note that due to the phase space contraction, 
$\nabla\cdot\mathbf{f}<0$, the transient-state dynamics
is always realized by stable limit-cycle or chaotic attractors.

\subsection{Bifurcation analysis of the symmetric network}

\begin{figure}[t]
\centering
\includegraphics[width=0.45\textwidth]
{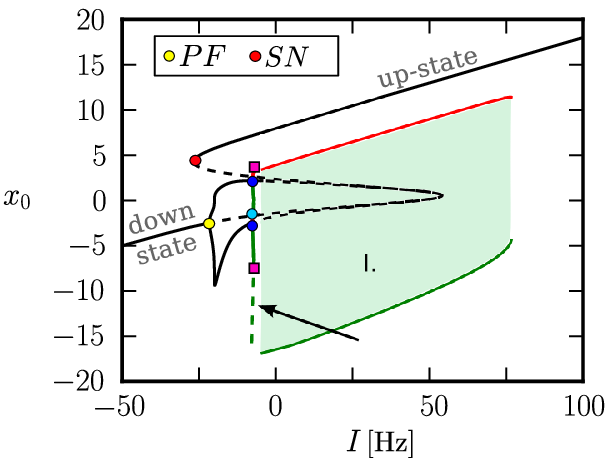}
\hskip 3ex
\includegraphics[width=0.45\textwidth]
{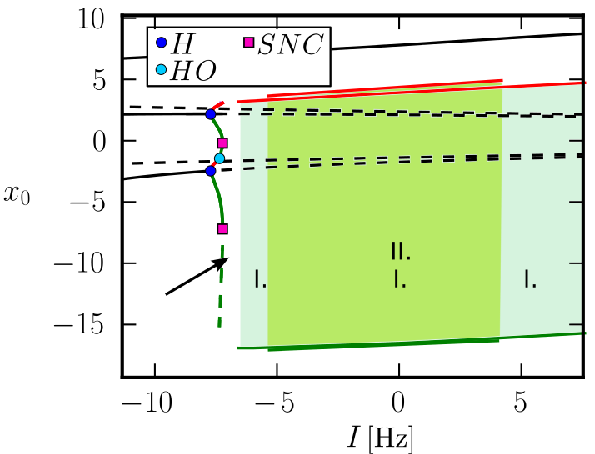}
\caption{\textit{Left:} Bifurcation diagram 
of the $N=4$ neuron network, projected on 
$x_0$, as a function of 
the input current $I$. \textit{Right:} A blow-up 
for the region with Hopf-bifurcations. Notations: 
solid black lines - stable fixpoints, 
dashed black lines - unstable fixpoints, 
red/green solid (dashed) lines - maximal/minimal 
amplitudes of stable (unstable) cycles, 
$PF$ - Pitchfork, $SN$ - saddle-node, 
$H$ - supercritical Hopf, $HO$ - homoclinic, 
and $SNC$ - saddle-node bifurcation of limit cycles. 
Different shades of green denote parameter intervals 
of class I.\ and II.\ transient-state dynamics. 
The arrow points to a chaotic window. 
The parameters are the same as for
Fig.~\ref{fig:timeseries_N=4}.}
\label{fig:bifurcations}
\end{figure}

We consider first the $N=4$ site symmetric network, 
as a function of the $I$, the global input, as it 
allows for a detailed bifurcation analysis of the 
system. Central sections of the bifurcation diagram,
created with PyDSTool \cite{clewley2012hybrid},
are presented in Fig.~\ref{fig:bifurcations}.

The symmetric network has $4$ stable fixpoints
in the absence of STSP ($\nu=0$), corresponding to 
active two-neuron cliques. Thanks to the $C_4$ symmetry, 
there exists a fifth, unstable state as well, for
which all four neurons have the same activity
$y^*_j\approx 0.3$, $j\in\{0,1,2,3\}$.  This symmetric 
solution splits up into three fixpoints for a large 
range of input values $I$ when STSP is turned on
($\nu=1$). This can be seen by solving 
graphically the fixpoint equation,
\begin{equation*}
\Gamma x^*= \left(2w_0 + z_0 u^*\varphi^*\right) y^*+I\,,
\quad
u^* = 1+(U_{\text{max}}-1)y^*\,,
\quad
\varphi^* = 1-u^*y^*/U_{\text{max}}
\end{equation*}
where $x^*_j=x^*$, $u^*_j=u^*$, $\varphi^*_j=\varphi^*$ 
and $y(x^*)=y^*$, for $j\in\{0,1,2,3\}$.
The two new fixpoints are generally referred 
to as up- and down-states \cite{cortes2013short}, 
with all neurons being nearly active $y_j\approx1$, 
or inactive $y_j\approx0$. This "S-shaped"
fixpoint solution forms the skeleton of the 
bifurcation diagram presented in
Fig.~\ref{fig:bifurcations}. The up-state 
is found to be stable in the considered 
parameter range even for certain negative inputs, 
being annihilated in the end by a saddle-node 
bifurcation ($SN$). The down-state splits,
on the other hand, into four clique-encoding 
solutions (seen as two branches, due to the $C4$ 
symmetry) via a higher order Pitchfork bifurcation 
($PF$), leaving the original state unstable.

For increasing input values of $I$ we find a cascade 
of bifurcations (akin to the series found in 
\cite{sandor2015versatile}), leading ultimately 
to transient-state dynamics type limit cycles. 
This cascade involves supercritical Hopf bifurcations 
($H$), saddle-node bifurcation of cycles ($SNC$), 
and homoclinic bifurcation $HO$.

We also found a narrow chaotic window (indicated by 
the arrow in Fig.~\ref{fig:bifurcations}), which we
however did not examine in detail.
The collapse of the chaotic attractor for larger 
inputs leads to class I.\ limit cycles (light green). 
Class II.\ travelling wave cycles (dark green) coexist 
with the flip-flop solutions, having very similar 
amplitudes (oscillating between the upper red and lower 
green lines) in the $x_0$ projection. As already discussed, 
the trajectories slow down, in both cases, close to the 
destabilized clique-fixpoints (dashed lines developing 
out of the Hopf points).

\begin{figure}[t]
\centering
\includegraphics[width=0.45\textwidth]
{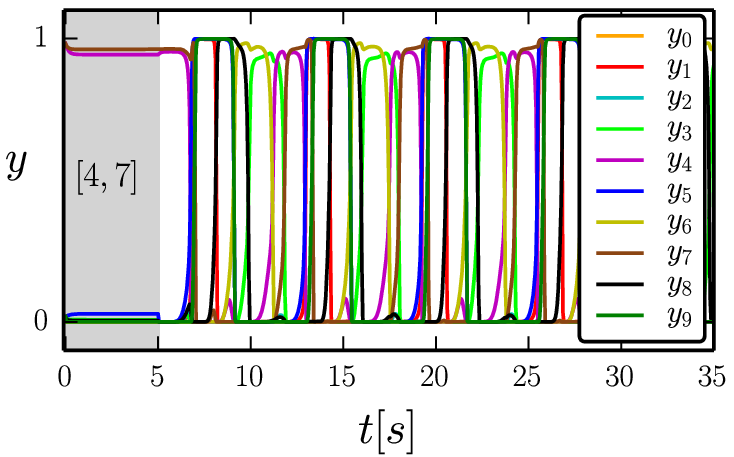}
\hskip 3ex
\includegraphics[width=0.45\textwidth]
{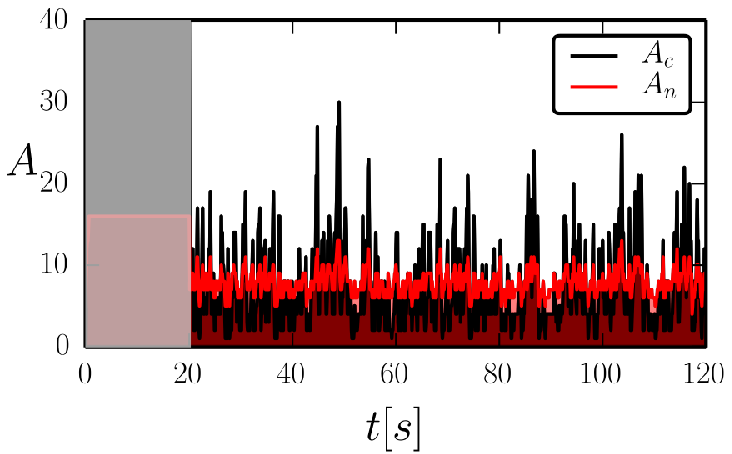}
\caption{\textit{Left}: Time-series of firing 
rates $y_k$ of the $N=10$ site random network.
\textit{Right}: Chaotic-like dynamics of the 
$N=100$ site network, showing the number of active 
neurons and cliques, $A_n$ and $A_c$ respectively, 
as a function of time.
STSP is turned off in the first $5/20$
seconds (\textit{left}/\textit{right}), as 
indicated by the respective grey-shaded areas,
with $A_c=1/41$ and $A_n=2/16$ initially 
(\textit{left}/\textit{right}).
The parameters are $I=0$, $a=0.4$, $\Gamma=10\,\text{s}^{-1}, 
$ $T_u=21\,\text{ms}$, 
$T_{\varphi}=706\,\text{ms}$, $\sigma_w=\sigma_z=10\,\text{Hz}$, 
together with $w_0=80\,\text{Hz}$, $z_0=-200\,\text{Hz}$ for the 
$N=10$ site, and $w_0=100\,\text{Hz}$, 
$z_0=-100\,\text{Hz}$ for the $N=100$ site networks.}
\label{fig:timeseries_N=10-100}
\end{figure}
 
\subsection{Complex activity patterns generated by random networks}

Considering larger networks, we show in
Fig.~\ref{fig:timeseries_N=10-100} that transient-state 
dynamics emerges generically out of fixpoint 
states when STSP is turned on. The sequence 
of activity patterns generated by STSP ($\nu=1$) 
then visits consecutively a subset of the  fixpoints
present at $\nu=0$, giving rise to multiple
coexisting limit cycles and/or chaotic attractors.

The time series of firing rates of the $N=10$ site network 
shown in Fig.~\ref{fig:timeseries_N=10-100} is, to
give an example:
\begin{center}
$[4,7]\rightarrow[1,5,7]\rightarrow[1,5,9]\rightarrow[5,8,9]
\rightarrow[6,8]\rightarrow[3,6]\rightarrow[3,4]\rightarrow\,,$
\end{center}
where we have identified cliques by the indices $k$ of the 
activated neurons (defined by $y_k>0.9$). For the selected 
parameters another limit cycle, 
$[1,5,7]\rightarrow[1,5,9]\rightarrow[5,8,9]\rightarrow[1,5,9]\rightarrow[1,5,7]$, 
with a complementary basin of attraction, in terms of initial states, 
exists. Defining with $A_n$ and $A_c$ the number of activated 
neurons and cliques respectively, we find that $A_n=2,3$ 
and $A_c=1$ during the cycle of the $N=10$ site
shown in Fig.~\ref{fig:timeseries_N=10-100}.

The number of co-activated cliques $A_c$ is controlled 
by the ratio of the average excitatory and inhibitory 
weights $w_0/|z_0|$, as shown by the data for the $N=100$
site network presented in Fig.~\ref{fig:timeseries_N=10-100}
(for which $w_0/|z_0|=1$). The initial state with 
turned off STSP corresponds to cell assemblies formed 
by multiple cliques, $A_c=\nobreak40$, with $A_n=16$ active neurons. 
The transient synaptic plasticity 
generates, furthermore, chaotic-like dynamics, with highly 
variable clique constellations, and  a average network 
activity of $\langle A_n/N\rangle\approx8\%$, a biologically 
realistic scenario.

\section{Conclusions}

We have investigated here the effect of a ubiquitous form of 
transient synaptic plasticity, STSP, on the dynamics of attractor 
neural networks. We find that STSP generically transforms stable 
attractors into attractor relics, inducing ongoing transient-state 
dynamics in terms of alternating sequences of active cell assemblies. 
We hence believe that it is essential to include STSP, which operates
on time scales of 0.1-1.0 seconds, in studies of biologically 
inspired neural networks. Applications range from 
working memory implementations \cite{mongillo2008synaptic}
to generation of motor control patterns by neural networks
\cite{cortes2013short, martin2016closed}.


\begin{footnotesize}

\end{footnotesize}


\end{document}